\documentclass{elsart}
\usepackage{amssymb, latexsym}
\usepackage[final]{graphics}

\newcommand{\eg}{e.g.}
\newcommand{\ie}{i.e.}
\newcommand{\ind}{\mathbf{1}}
\newcommand{\interior}[1]{{\rm int}\,#1}
\newcommand{\mathfrac}[2]{\mbox{$\frac{#1}{#2}$}}

\begin{document}


\begin{frontmatter}

\title{Quasi Markovian behavior in mixing maps}
\author{Brian R. La Cour\thanksref{Email}} and
\author{William C. Schieve}
\address{Ilya Prigogine Center for Studies in Statistical Mechanics and Complex Systems \\
         The University of Texas at Austin \\
         Austin, Texas 78712}
\thanks[Email]{Corresponding author. E-mail address: blacour@physics.utexas.edu}


\begin{abstract}
We consider the time dependent probability distribution of a coarse grained observable $Y$ whose evolution is governed by a discrete time map.  If the map is mixing, the time dependent one-step transition probabilities converge in the long time limit to yield an ergodic stochastic matrix.  The stationary distribution of this matrix is identical to the asymptotic distribution of $Y$ under the exact dynamics.  The $n$th time iterate of the baker map is explicitly computed and used to compare the time evolution of the occupation probabilities with those of the approximating Markov chain.  The convergence is found to be at least exponentially fast for all rectangular partitions with Lebesgue measure.  In particular, uniform rectangles form a Markov partition for which we find exact agreement.
\end{abstract}

\begin{keyword}
master equation; mixing; Markov chain; baker map; chaos
\PACS{2.50.Ga; 05.20.-y; 05.40.+j; 05.45.+b; 05.70.Ln}
\end{keyword}

\end{frontmatter}


\section{Introduction}
\label{sect:Introduction}

The technique of coarse graining has often been used to obtain irreversible behavior from systems which, though they may be reversible, are ``sufficiently'' chaotic \cite{Ehrenfest1959,Kampen1962,Misra1980}.  The procedure is to describe the time evolution of the system in terms of macroscopic variables whose preimages form the cells of the partition.  More abstractly, coarse graining may be described as a contraction of the exact dynamics onto a corresponding symbolic dynamics \cite{Antoniou1997}.  One of the goals of this approach is be able to describe the macroscopic dynamics in a self-contained form, in other words, to obtain macroscopic laws which do not refer to the underlying microstates.  This may be possible if the macroscopic dynamics satisfy a master equation \cite{Kampen1981}.

  It has long been realized, however, that the evolution of observables under reversible microdynamics are generally non-Markovian due to memory effects \cite{Kampen1962}.  While it may be possible to construct a partition for which the resulting symbolic dynamics are exactly Markovian \cite{Misra1980,Nicolis1988}, such partitions are exceptional and in general do not correspond to the macroscopic variables one is interested in.  Assumptions of molecular chaos have in the past been used to argue that the Markov assumption is well approximated for complex macroscopic systems \cite{Kampen1962}; however, it is unclear whether such assumptions would be consistent with the underlying microscopic dynamics.  What are needed are sufficient conditions on the dynamics, as well as the macroscopic observables, to ensure that the Markov assumption is well approximated.  This has been done, {\eg}, for quantum systems weakly interacting with a heat bath \cite{Davies1976}, but it is not clear how those results may apply in more general settings.  Since the master equation is characterized by only short-term memory effects, it is natural to suppose that decaying correlations should at least be a necessary condition.  Thus, one is led to consider systems which are mixing.

  We begin in Section~\ref{sect:Generalized Master Equation} with a discussion of the general problem for discrete time maps.  An equation for the time dependent occupation probabilities, {\ie} the generalized master equation, is derived.  (Since it involves one-step transition probabilities that are in general dependent upon both the initial cell occupied and the iteration time, the generalized master equation is not generally a true master equation.)  In Section~\ref{sect:Mixing and Approximating Markov Chains} we consider the asymptotic behavior of these transition probabilities for long times.  For mixing maps it is shown that the one-step transition probabilities converge to an ergodic stochastic matrix.  The corresponding Markov chain is found to converge to the same stationary distribution as that of the exact process.  The baker map is considered in Section~\ref{sect:The Baker Map} as an exactly solvable system which nevertheless contains many of the important features of more physical systems.  An explicit expression is obtained for the $n$th iterate of a point under this map, thus allowing us to compute exact time dependent occupation probabilities.  For rectangular cells with Lebesgue measure it is shown that these probabilities converge at least exponentially fast.  In Section~\ref{sect:Comparison of Exact and Approximating Processes} we compare the occupation and transition probabilities determined by the baker map with their corresponding approximations using the approximating Markov chain.  In particular it is shown that for the baker map any uniform rectangular partition is a Markov partition for which the Markov approximation is exact.  The main conclusions are summarized and discussed in Section~\ref{sect:Discussion}.


\section{Generalized Master Equation}
\label{sect:Generalized Master Equation}

Let $(X,{\mathcal A},P)$ be a probability space and let $\phi$ be a
measurable map on $X$.  We assume that $P$ has a density $\rho$ with respect
to some finite measure $\mu$, {\ie} $P\prec\mu$, and that $\phi$ is
$\mu$-measure preserving, {\ie} $\mu\circ\phi^{-1}=\mu$.  For $x_0\in X$ and
$n\in\Zset$, $\phi^n(x_0)$ is the $n^{\rm th}$ iterate of $x_0$ under
$\phi$.  (If $\phi$ is invertible, then $\phi^n=(\phi^{-1})^{|n|}$ for
$n<0$; otherwise, $\phi^n$ is undefined.)  The preimage of a measurable set
$A$ under $\phi^n$ will be denoted by $\phi^{-n}(A)$ and represents the set
of points that map into $A$ after $n$ time steps.  Thus, $P[\phi^{-n}(A)]$
is the probability that an iterate will be in $A$ at time $n$.

We wish to consider the evolution of a phase function $Y$ of the form
\begin{equation}
Y(x) \equiv \sum_{i\in I} y_i \ind_{C_i}(x)~,
\end{equation}
where $I$ is a finite index set, $\{y_i\}_{i\in I}$ are distinct real
numbers, and $\ind_{C_i}$ is the indicator (characteristic) function on
$C_i\in{\mathcal A}$.  Without loss of generality, we may assume
$\{C_i\}_{i\in I}$ is a disjoint partition of $X$.  We further suppose that
the partition is nontrivial in the sense that $\mu[C_i]>0$ for all $i\in I$.
Let $Y_n \equiv Y\circ\phi^n$ and define the transition probabilities
$T_{i,j}^{(n)}$ and $T_{i,k,j}^{(m,n)}$ as follows:
\begin{eqnarray}
T_{i,j}^{(n)} & \equiv & P[\phi^{-n}(C_j)~|~C_i]~,
\label{eqn:TPG1S}\\ 
T_{i,k,j}^{(m,n)} & \equiv & P[\phi^{-n}(C_j)~|~\phi^{-m}(C_k) \cap C_i]~,
\label{eqn:TPG2S} 
\end{eqnarray}
for $P[C_i]>0$ and $P[\phi^{-m}(C_k) \cap C_i]>0$, respectively, and zero
otherwise.  We see that $T_{i,j}^{(n)}$ is the probability that $Y_n=y_j$
given that $Y_0=y_i$, and $T_{i,k,j}^{(m,n)}$ is the probability that
$Y_n=y_j$ given that $Y_m=y_k$ and $Y_0=y_i$.  In general,
$T_{i,k,j}^{(m,n)}$ will depend upon $i$, so that $\{Y_n\}_{n\in\Zset}$ is
generally a non-Markovian random process.

Using Eqs.~(\ref{eqn:TPG1S}) and (\ref{eqn:TPG2S}) we obtain the generalized Chapman-Kolmogorov equation for this process:
\begin{equation}
T_{ij}^{(n+m)} = \sum_k T_{ik}^{(n)} T_{ikj}^{(n,n+m)}~.
\label{eqn:GCK} 
\end{equation}
Let $W_{ikj}^{(n)} \equiv T_{ikj}^{(n,n+1)}$ denote the one-step transition probability at time $n$ conditioned on $C_i$.  We see that $W_{ikj}^{(n)}$ is the probability that an iterate in cell $C_k$ at time $n$ will make a transition to cell $C_j$ at the next time step, given that it started in cell $C_i$.  Using Eq.~(\ref{eqn:GCK}) it can be shown by induction over $n$ that
\begin{equation}
T_{ij}^{(n)} = \sum_{k_1} \cdots \sum_{k_{n-1}} \prod_{m=1}^n W_{i,k_{m-1},k_m}^{(m-1)}~,
\label{eqn:GMP} 
\end{equation}
where $k_0=i$ and $k_n=j$.  Let $p_j(n) \equiv P[\phi^{-n}(C_j)]$ denote the
occupation probability of cell $C_j$ at time $n$ and note that
\begin{equation}
p_j(n)=\sum_i p_i(0) T_{ij}^{(n)}~.
\label{eqn:GME} 
\end{equation}
This is the generalized master equation for the exact occupation probabilities.  In general, $W_{ikj}^{(n)}$ depends on both $i$ and $n$, so Eq.~(\ref{eqn:GME}) contains memory effects, {\ie} the process is not Markovian.  Nevertheless we might expect quasi Markovian behavior for maps with decaying correlations.  This suggests that mixing may be a key property to consider.


\section{Mixing and Approximating Markov Chains}
\label{sect:Mixing and Approximating Markov Chains}

The map $\phi$ defined in Sect.~\ref{sect:Generalized Master Equation} will
be \emph{mixing} with respect to $\mu$ \cite{Lasota1994} if for all
measurable sets $A$ and $B$,
\begin{equation}
\lim_{n\rightarrow\infty}\mu[\phi^{-n}(A) \cap B] = P_*[A] \mu[B]~,
\label{eqn:MD} 
\end{equation}
where $P_* \equiv \frac{1}{\mu[X]}\mu$.  From Eq.~(\ref{eqn:MD}) it follows \cite[p.~72]{Lasota1994} that if $P\prec\mu$ and $P[B]>0$ then
\begin{equation}
\lim_{n\rightarrow\infty}P[\phi^{-n}(A)~|~B]=P_*[A].
\label{eqn:MC} 
\end{equation}

Using Eq.~(\ref{eqn:MC}) we find that for $P[C_i]>0$,
\begin{equation}
\lim_{n\rightarrow\infty}T_{i,j}^{(n)}=
\lim_{n\rightarrow\infty}P[\phi^{-n}(C_j)~|~C_i] = P_*[C_j]~,
\label{eqn:AP} 
\end{equation}
the limit being zero otherwise.  Note that from Eqs.~(\ref{eqn:GME}) and
(\ref{eqn:AP}) it follows that $p_j(n)\rightarrow P_*[C_j]$ as
$n\rightarrow\infty$.  Thus, mixing implies an asymptotic convergence of
solutions to Eq.~(\ref{eqn:GME}) for any initial $P\prec\mu$.  From
Eq.~(\ref{eqn:MC}) we also note that if $P[C_i]>0$ then
$P[\phi^{-n}(C_j)\cap C_i]>0$ for all $n$ sufficiently large.  (Recall that
$P_*[C_j]>0$ for a nontrivial partition.)  Thus for $P[C_i]>0$ we similarly
have
\[ \lim_{n\rightarrow\infty}T_{i,k,j}^{(n,n+m)} =
\lim_{n\rightarrow\infty}\frac{P[\phi^{-n}(\phi^{-m}(C_j) \cap
C_k)~|~C_i]}{P[\phi^{-n}(C_k)~|~C_i]} = P_*[\phi^{-m}(C_j)~|~C_k]~. \]
The one-step transition probabilities therefore have the following asymptotic form, independent of $k$:
\begin{equation}
\lim_{n\rightarrow\infty} W_{k,i,j}^{(n)} = P_*[\phi^{-1}(C_j)~|~C_i] \equiv
W_{i,j}~.
\label{eqn:AW}
\end{equation}

Since $W$ is clearly a stochastic matrix, it must correspond to some Markov chain $\{\tilde{Y}_n\}_{n\in\Zset}$ on the state space $\{y_i\}_{i\in I}$.  We shall refer to $\{\tilde{Y}_n\}_{n\in\Zset}$ as the \emph{approximating Markov chain} for the exact process $\{Y_n\}_{n\in\Zset}$.  Let $q_i(n)$ denote the probability that $\tilde{Y}_n=y_i$ for $i\in I$.  In contrast to Eq.~(\ref{eqn:GME}), we have simply \cite[p.~95]{Kampen1981}
\begin{equation}
q_j(n)=\sum_i q_i(0) (W^n)_{i,j}~.
\label{eqn:AME} 
\end{equation}
The condition $T_{i,j}^{(n)}=(W^n)_{i,j}$ is thus seen to be the necessary and sufficient condition for exact agreement between $\{Y_n\}$ and $\{\tilde{Y}_n\}$, assuming $p_j(0)=q_j(0)$ \cite{Nicolis1988}.  The following theorem establishes that, even without this equality, the two processes converge to the same stationary distribution.


\begin{thm}
If $\phi$ is mixing with respect to a finite invariant measure $\mu$ and
$\{C_i\}_{i\in I}$ is a nontrivial finite partition of $X$, then any
approximating Markov chain is ergodic.
\end{thm}
\begin{pf}
A Markov chain is ergodic if it is irreducible, aperiodic, and possesses a
stationary distribution \cite[p.~109]{Ross1983}.  Let $i,j\in I$ be fixed
and take $P=P_*$ in Eq.~(\ref{eqn:GMP}).  Given that $\phi$ is mixing, we
know $T_{i,j}^{(n)}\rightarrow P_*[C_j]$.  Since $\mu[C_j]>0$ by assumption,
$P_*[C_j]>0$ and so $T_{i,j}^{(n)}>0$ for all $n>N_{i,j}$, where $N_{i,j}$
is some positive integer.  Thus,
\begin{eqnarray*}
0 < T_{i,j}^{(n)}
&=&   \sum_{k_1,\ldots,k_{n-1}\in K_n} \prod_{m=1}^n P_*[\phi^{-m}(C_{k_m})~|~\phi^{-(m-1)}(C_{k_{m-1}})\cap C_i] \\
&\le& \frac{1}{p_n} \sum_{k_1,\ldots,k_{n-1}\in K_n} \prod_{m=1}^n P_*[\phi^{-m}(C_{k_m})\cap \phi^{-(m-1)}(C_{k_{m-1}})\cap C_i]~,
\end{eqnarray*}
where $k_1=i$, $k_n=j$, and
\begin{eqnarray}
K_n &=& \left\{k_1,\ldots,k_{n-1}\in I: \prod_{m=1}^n
P_*[\phi^{-(m-1)}(C_{k_{m-1}})\cap C_i] > 0 \right\}~,\\ p_n &=&
\min_{k_1,\ldots,k_{n-1}\in K_n}\left\{\prod_{m=1}^n
P_*[\phi^{-(m-1)}(C_{k_{m-1}})\cap C_i]\right\}~.
\end{eqnarray}
We are assured that $K_n$ is nonempty since, were it empty, $T_{i,j}^{(n)}$ would be zero.  Since we have taken $n$ to be sufficiently large so that $T_{i,j}^{(n)}>0$, it follows that $K_n$ is indeed nonempty, and so $p_n>0$.  Multiplying both sides of the inequality by $p_n$ we conclude that
\[ \sum_{k_1,\ldots,k_{n-1}\in K_n} \prod_{m=1}^n P_*[\phi^{-m}(C_{k_m})\cap
\phi^{-(m-1)}(C_{k_{m-1}})\cap C_i] > 0~. \]
Now, since $W_{i,j}\ge P_*[\phi^{-1}(C_j)\cap C_i] $ and $P_*$ is invariant,
\begin{eqnarray*}
(W^n)_{i,j}
& = & \sum_{k_1}\cdots\sum_{k_{n-1}} \prod_{m=1}^n W_{k_{m-1}, k_m} \\
&\ge& \sum_{k_1}\cdots\sum_{k_{n-1}} \prod_{m=1}^n P_*[\phi^{-1}(C_{k_m})\cap C_{k_{m-1}}] \\
&=&   \sum_{k_1}\cdots\sum_{k_{n-1}} \prod_{m=1}^n P_*[\phi^{-m}(C_{k_m})\cap \phi^{-(m-1)}(C_{k_{m-1}})] \\
&\ge& \sum_{k_1,\ldots,k_{n-1}\in K_n} \prod_{m=1}^n P_*[\phi^{-m}(C_{k_m})\cap \phi^{-(m-1)}(C_{k_{m-1}})\cap C_i]~.
\end{eqnarray*}
It then follows that $(W^n)_{i,j}>0$ for $n>N_{i,j}$.  Since $i$ and $j$ were arbitrary, this proves irreducibility.

A Markov chain is aperiodic if it has period $d=1$.  If we suppose that $d>1$, however, then for $1\le r<d$ we would have that $(W^{md+r})_{i,i}=0$ for all $m\in\Nset$.  We have just shown, however, that $(W^n)_{i,i}>0$ for all $n$ sufficiently large.  Since $d$ cannot be smaller than 1, we conclude that $d=1$.

Finally, it may be readily verified that $q_i(0)=P_*[C_i]$ for $i\in I$ is a stationary state of $W$.
\qed
\end{pf}


An ergodic Markov chain will converge to a unique stationary distribution.  We thus conclude that for all $i,j$
\begin{equation}
\lim_{n\rightarrow\infty}(W^n)_{i,j} = P_*[C_j]~.
\end{equation}
The limit is both geometric in rate and uniform in $i, j$ \cite[p.~124]{Cox&Miller1965}.  We conclude that both $q_j(n)$ and $p_j(n)$ approach the same asymptotic value of $P_*[C_j]$.  The asymptotic distribution will be uniform whenever the partition is uniform, {\ie} when $\mu[C_i]=\mu[C_j]$ for all $i,j\in I$, and the following theorem gives the necessary and sufficient conditions for this to be true.


\begin{thm}
The partition is uniform if and only if $W$ is doubly stochastic.
\end{thm}
\begin{pf}
If the partition is uniform then \[ \sum_i W_{i,j} = \frac{1}{\mu[C_1]}
\sum_i \mu[\phi^{-1}(C_j)\cap C_i] \\ = \frac{1}{\mu[C_1]}
\mu[\phi^{-1}(C_j)] = 1~. \] Now suppose $W$ is doubly stochastic and note
that for any $n\in\Nset$, $W^n$ will also be doubly stochastic.  Since the
partition is finite, \[ 1 = \lim_{n\rightarrow\infty} \sum_i (W^n)_{i,j} =
\sum_i \lim_{n\rightarrow\infty} (W^n)_{i,j} = \sum_i P_*[C_j]~. \] If there
are $M$ partition cells, then $\mu[C_j]=\frac{1}{M}\mu[X]$ for all $j\in I$.
\qed
\end{pf}


\section{The Baker Map}
\label{sect:The Baker Map}

  We now consider the above results in light of a particular example, the baker map, which is defined on $X=[0,1)^2$ as follows \cite{Lasota1994}:
\begin{equation}
\phi(x,y) = \left(2x,~\frac{y}{2}\right) \ind_{[0,1/2)}(x) +
\left(2x-1,~\frac{y}{2}+\frac{1}{2}\right) \ind_{[1/2,1)}(x)~.
\label{eqn:BM} 
\end{equation}
The baker map is a Lebesgue measure preserving map of the Bernoulli type and
hence is mixing \cite{Arnold&Avez1968}.  It is also time reversal invariant
in the sense that $\phi\circ R=R\circ\phi^{-1}$ for the transformation
$R(x,y)=(y,x)$.  Thus, it possesses all the salient features of a closed
Hamiltonian system with time reversal invariance while permitting a simple
and exact analysis.  An example of an real-world optical system which
produces a baker transformation is described in \cite{Hannay1994}.

We wish to explicity compute the probability $p_j(n)$ defined in Section
\ref{sect:Generalized Master Equation}.  To do so we will need to calculate
probabilities using the given initial probability density.  Now, given a
probability density $\rho$, the density after $n$ time steps is given by
$U^n\rho$, where $U$ is the Perron-Frobenius operator.  Although the
generalized spectral decomposition of $U$ (and hence $U^n$) for the baker
map has been constructed \cite{Hasegawa1992}, it cannot generally be used
for computing probabilities.  However, since $\phi$ is invertible and area
preserving, $U^n\rho=\rho\circ\phi^{-n}$ \cite[p.~47]{Lasota1994}.
Furthermore, since $\phi^{-n}=R\circ\phi^n\circ R$, to compute $U^n$ for
$n\in\Zset$ it suffices to determine $\phi^n$ for $n\in\Nset$.

\begin{thm}
Let $\phi$ denote the baker map and let $U$ be the corresponding
Perron-Frobenius operator.  For any integer $n\in\Nset$ and point
$(x,y)\in[0,1)^2$,
\begin{equation}
\phi^n(x,y) = \sum_{m=1}^{2^n}
\left(2^nx-m+1,~\frac{y}{2^{n}}+\alpha(m)\right)
\ind_{\left[\frac{m-1}{2^n},\frac{m}{2^n}\right)}(x)~,
\label{eqn:TIBM} 
\end{equation}
where
\begin{equation}
\alpha(m) \equiv
\left\{
\begin{array}{ll}
0, & \mbox{if $m=1$},\\
\alpha(m-2^{n-1})+1/2^n, & \mbox{if $m\in\{2^{n-1}+1, \ldots, 2^n\}$~.}
\end{array}
\right.
\label{eqn:AC} 
\end{equation}
\end{thm}
\begin{pf}
The proof is by induction on $n$.  For $n=1$ Eq.~(\ref{eqn:TIBM}) clearly
holds.  By Eq.~(\ref{eqn:BM}) we have, for
$\phi^n(x,y)=\phi(\phi^{n-1}(x,y))$,
\begin{eqnarray*}
\phi^n(x,y)
&=& \sum_{m=1}^{2^{n-1}} \left(2^{n}x-m+1,~\frac{y}{2^{n}}+\alpha(m)\right) \ind_{\left[\frac{m-1}{2^{n-1}},\frac{m}{2^{n-1}}\right)}(2x) ~ \ind_{[0,1/2)}(x) \\
& & + \sum_{m=1}^{2^{n-1}} \left(2^nx-2^{n-1}-m+1,~\frac{y}{2^n}+\frac{1}{2^n}+\alpha(m)\right) \\
& & \times~ \ind_{\left[\frac{m-1}{2^{n-1}},\frac{m}{2^{n-1}}\right)}(2x-1) ~ \ind_{[1/2,1)}(x)
\end{eqnarray*}
In the second sum we make the substitution $m=m'-2^{n-1}$ and note that $\frac{1}{2^n}+\alpha(m)=\alpha_n(m')$, by Eq.~(\ref{eqn:AC}).  Finally, we note the following identities:
\[
\ind_{\left[\frac{m-1}{2^{n-1}},\frac{m}{2^{n-1}}\right)}(2x) ~ \ind_{[0,1/2)}(x) = \ind_{[\frac{m-1}{2^n},\frac{m}{2^n})}(x),
\]
since $m\le 2^{n-1}$ is equivalent to $m/2^n \le \half$, and
\[
\ind_{\left[\frac{m-1}{2^{n-1}},\frac{m}{2^{n-1}}\right)}(2x-1) ~ \ind_{[1/2,1)}(x) = \ind_{[\frac{m'-1}{2^n},\frac{m'}{2^n})}(x),
\]
since $m' \ge 2^{n-1}+1$ is equivalent to $(m'-1)/2^n \ge \half$.  Thus,
\begin{eqnarray*}
\phi^n(x,y)
&=& \sum_{m=1}^{2^{n-1}} \left(2^nx-m+1,~\frac{y}{2^n}+\alpha(m)\right) \ind_{\left[\frac{m-1}{2^n},\frac{m}{2^n}\right)}(x) \\
& & + \sum_{m'=2^{n-1}+1}^{2^n} \left(2^nx-m'+1, ~\frac{y}{2^n}+\alpha(m')\right) \ind_{\left[\frac{m'-1}{2^n},\frac{m'}{2^n}\right)}(x),
\end{eqnarray*}
which is identical to Eq.~(\ref{eqn:TIBM}).
\qed
\end{pf}

Note that the sum in Eq.~(\ref{eqn:TIBM}) has only one nonzero term, namely, $m=m_n(x) \equiv \lfloor 2^nx \rfloor+1$.  Thus it may be written more compactly as
\begin{equation}
\phi^n(x,y) = \left( 2^nx\bmod1,~ \frac{y}{2^n}+\alpha(m_n(x)) \right)~.
\end{equation}

Using Eq.~(\ref{eqn:TIBM}) and the fact that
$U^n\rho=\rho\circ(R\circ\phi^n\circ R)$, we find that for a rectangular
region $C=[a,b)\times[c,d)$,
\begin{equation}
P[\phi^{-n}(C)] =\sum_{l=m_n(c)}^{m_n(d)} ~
\int_{\gamma_{n,l}(c)}^{\delta_{n,l}(d)} ~
\int_{\alpha(l)+a/2^n}^{\alpha(l)+b/2^n} \rho(u,v)~\d u~\d v~,
\label{eqn:OP} 
\end{equation}
where $\gamma_{n,l}(c)=\max\{2^nc,l-1\}-l+1$ and $\delta_{n,l}(d)=\min\{2^nd,l\}-l+1$.  (We define $m_n(1) \equiv 2^n$.)  Note that $\gamma_{n,l}(c)=2^nc\bmod1$ for $l=m_n(c)$ and is zero otherwise, while $\delta_{n,l}(d)=2^nd\bmod1$ for $l=m_n(d)$ and is one otherwise.

Let us now consider an initial density of the form
\begin{equation}
\rho(x,y) \equiv \sum_i p_i(0) / \mu_L[C_i]~\ind_{C_i}(x,y),
\label{eqn:CGD} 
\end{equation}
where the $p_i(0)$'s are given constants and $\mu_L[C_i]$ is the Lebesgue measure of cell $C_i$.  (Of course, such a density may be obtained by pre-coarse graining an arbitrary $\rho$, in which case $p_i(0) = \int_{C_i}\rho(x,y)~\d x\d y$.)  If the cells are rectangular with $C_i=[a_i,b_i)\times[c_i,d_i)$ then
\begin{equation}
T_{i,j}^{(n)} = \mu_L[\phi^{-n}(C_j)~|~C_i] = \sum_{l=m_n(c_j)}^{m_n(d_j)}
\frac{f_n[\alpha(l)]~g_{n,l}}{(b_i-a_i)(d_i-c_i)}~,
\label{eqn:TPUP} 
\end{equation}
where
\begin{eqnarray}
f_n(\alpha) & \equiv & \vartheta(\min\{b_i,\alpha+b_j/2^n\}-\max\{a_i,\alpha+a_j/2^n\})~,\\
g_{n,l}     & \equiv & \vartheta(\min\{d_i,\delta_{n,l}(d_j)\}-\max\{c_i,\gamma_{n,l}(c_j)\})~.
\end{eqnarray}
(The function $\vartheta$ is defined such that $\vartheta(x)=x$ for $x>0$ and $\vartheta(x)=0$ otherwise.)  The occupation probability $p_j(n)$ is now found using Eq.~(\ref{eqn:GME}).

As an example, consider $\rho=\frac{1}{\mu_L[C_j]} \ind_{C_j}$.  For $n<\log_2(1/d_j)$ we have
\begin{equation}
p_j(n) = \frac{1}{2^n}\frac{\vartheta(b_j-2^na_j) \vartheta(d_j-2^nc_j)}{(b_j-a_j)(d_j-c_j)} \le \frac{1}{2^n}~.
\end{equation}
If $a_j=c_j=0$, we have an initially exponential decay of  $p_j(n)$ before it converges to the asymptotic value $\mu_L[C_j]=b_jd_j$.

This example suggests a general convergence rate which is at least exponentially fast.  Let us consider the rate of convergence of $T_{i,j}^{(n)}$.  Assuming $n>\log_2[(b_j-a_j)/(b_i-a_i)]$ and $n\gg\log_2(d_j-c_j)^{-1}$, we have
\[
\mu_L[C_i]~T_{i,j}^{(n)}
= f_n[\alpha(l_c)] ~ g_{n,l_c} + f[\alpha(l_d)] ~ g_{n,l_d} + \sum_{l=l_c+1}^{l_d-1} f_n[\alpha(l)] ~,
\]
where $l_c=m_n(c_j)$ and $l_d=m_n(d_j)$.  The sequence $\alpha(1), \alpha(2), \ldots$ is dense in $[0,1)$, taking values which are uniformly distributed throughout the interval.  Thus, we may approximate the sum by an integral for large $n$.
\[
\sum_{l=l_c+1}^{l_d-1} f_n[\alpha(l)] \approx 2^n(d_i-c_i)(d_j-c_j)\int_0^1 f_n(\alpha) \d\alpha~,
\]
where we note that
\begin{equation}
f_n(\alpha)
=\left\{
\begin{array}{ll}
\alpha-a_i+b_j/2^n, & \mbox{if $a_i-\frac{b_j}{2^n}\le\alpha<a_i-\frac{a_j}{2^n}$,}\\
(b_j-a_j)/2^n       & \mbox{if $a_i-\frac{a_j}{2^n}\le\alpha<b_i-\frac{b_j}{2^n}$,}\\
b_i-a_j/2^n-\alpha  & \mbox{if $b_i-\frac{b_j}{2^n}\le\alpha<b_i-\frac{a_j}{2^n}$,}\\
0~,                 & \mbox{otherwise}~.
\end{array}
\right.
\end{equation}
Upon integrating we find that
\[
\left|T_{i,j}^{(n)}-\mu_L[C_j]\right|
\approx \frac{f_n[\alpha(l_c)]~g_{n,l_c}+f_n[\alpha(l_d)]~g_{n,l_d}}{\mu_L[C_i]}
\le \frac{2(b_j-a_j)}{2^n \mu_L[C_i]}~.
\]
Thus we have convergence which is at least exponentially fast.  (Recall that convergence for the approximating Markov chain is always exponential.)


\section{Comparison of Exact and Approximating Processes}
\label{sect:Comparison of Exact and Approximating Processes}

We have found that for mixing maps both $p_j(n)$ and $q_j(n)$ converge to the same limit, $P_*[C_j]$, as $n\rightarrow\infty$.  If furthermore $p_j(0)=q_j(0)$ for all $j\in I$, then we expect only transient deviations of $p_j(n)$ from its Markovian counterpart $q_j(n)$.  The difference between these two probabilities may be measured by an average over all cells of the $\ell^{\infty}$ norms:
\begin{equation}
D_p \equiv \frac{1}{M} \sum_{j=1}^M \sup_{n\in\Nset} \left| p_j(n)-q_j(n) \right|,
\end{equation}
where $M$ is the total number of cells.  Note that $0\le D_p\le 1$, the value of $D_p$ depending upon both the initial measure $P$ and the partition chosen.  In particular, if $P=P_*$ then $p_j(n)=P_*[C_j]=q_j(n)$ and $D_p=0$.

We now return to the baker map and consider a partition of two cells: $C_1=[0,s)\times[0,1)$ and $C_2=[s,1)\times[0,1)$.  For $s\le\half$, the transition matrix given by Eq.~(\ref{eqn:AW}) is
\begin{equation}
W=\left(
\begin{array}{cc}
\half & \half \\
\frac{s}{2(1-s)} & \frac{2-3s}{2(1-s)}
\end{array}
\right)~.
\end{equation}
The transition matrix for $s>\half$ is obtained by interchanging $s$ with $1-s$ and taking the transpose.

\begin{figure}
\resizebox{100mm}{!}{\includegraphics*[40mm,40mm][150mm,115mm]{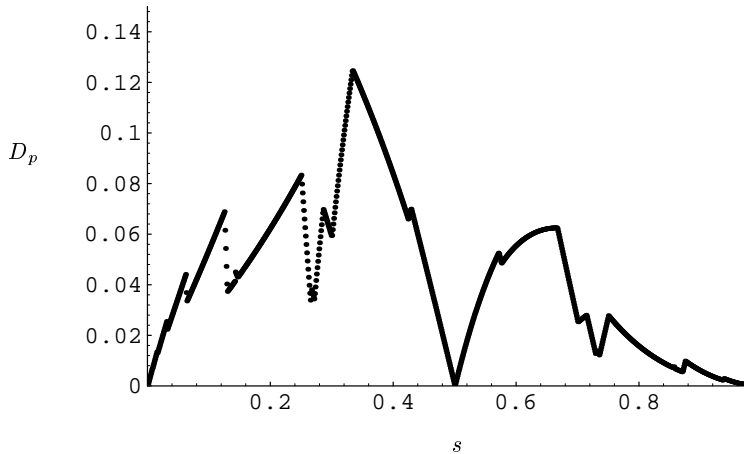}}
\caption{The above plot shows the values of $D_p$ for a partition of two
vertical cells of width $s$ and $1-s$.  Note that $D_p$ goes to zero near
$s=0,\half,1$.}
\label{fig:Dpvs} 
\end{figure}

In Fig.~\ref{fig:Dpvs} we display the average deviation $D_p$ for $0<s<1$.  The asymmetry in the graph is due to the fact that the initial measure is on $C_1$; a mirror image of it is obtained if $C_2$ is used as the starting cell.  For $s$ near 0, $\half$, or 1 we note that $D_p$ goes to zero, implying exact agreement between $p_j(n)$ and $q_j(n)$ for all $j$ and $n$.  This can be understood as a consequence of the following theorem:
\begin{thm}
Let $R_{i,j}=[\frac{i-1}{M},\frac{i}{M}]\times[\frac{j-1}{N},\frac{j}{M}]$ for $i\in\{1,\ldots,M\}$, $j\in\{1,\ldots,N\}$ and $M,N\in\Nset$.  Then $\{R_{i,j}\}$ is a Markov partition for the baker map.
\label{thm:MPBM} 
\end{thm}
A proof of this theorem is given in the appendix.  Since the initial measure was taken to be uniform with respect to the maximum entropy measure $\mu_L$, it follows that $\{Y_n\}$ is a true Markov chain \cite{Cornfeld1982}.

\section{Discussion}
\label{sect:Discussion}

We have shown that for mixing maps the time dependent occupation and one-step transition probabilities, $p_j(n)$ and $W_{kij}^{(n)}$ respectively, converge to values which are independent of both the index $k$ and the initial probability density $\rho$.  The resulting asymptotic transition probabilities $W_{ij}$ form an ergodic stochastic matrix whose unique asymptotic distribution is identical to that of the exact, non-Markovian, process.

To examine the detailed evolution of $p_j(n)$ we considered the baker map
$\phi$.  An explicit expression for $\phi^n(x,y)$ was found with which we
were able to determine explicit formulas for $p_j(n)$ in the case of an
initial density which is uniform on rectangular cells.  It was shown that in
this case the occupation probability $p_j(n)$ converges at least
exponentially fast, {\ie} no slower than $q_j(n)$.

Finally, we studied the average deviation between $p_j(n)$ and $q_j(n)$, which we denoted by $D_p$.  Partitions of two vertical rectangular cells were considered for various cell widths.  It was found that agreement was best when all cells were similar, but overall agreement was generally good ($D_p<0.15$).  In particular, uniform rectangular partitions gave exact agreement ($D_p=0$).  This was understood to be a consequence of the fact that such partitions are in fact Markov partitions for the baker map.  In general, however, the agreement between the exact and approximating processes will depend upon how strongly the system is mixing.  Since the baker map is not only mixing but also a Bernoulli system, it is not clear whether such good agreement is typical of most mixing systems.

While the time evolution of a coarse grained observable under a deterministic map may not be Markovian, for mixing maps this evolution is at least approximately Markovian in the sense that both $\{Y_n\}$ and $\{\tilde{Y}_n\}$ have the same distribution initially and in the distant future.  (If the map is invertible then they also have the same distribution in the distant past.)  Thus, the two processes do not differ qualitatively and in some cases agree quite well quantitatively.

We have not addressed the problem of obtaining closed form deterministic laws for macroscopic variables.  This may be possible for ``many to one'' observables corresponding to strongly nonuniform partitions.  In general, however, such macroscopic laws are expected to be only approximately deterministic, as befits any realistic model.


\begin{ack}
This research was supported in part by a grant from the U.S. Department of Energy (Grant No. DE-FG03-94ER14465).
\end{ack}


\appendix
\section*{Appendix}

Here we prove the claim that every uniform rectangular partition is a Markov partition for the baker map.  Note that this is not a true partition since the boundaries of the cells overlap; however, the boundaries are of measure zero and may easily be modified to form a true partition.
\begin{pf} (Theorem \ref{thm:MPBM})
The collection $\{R_{i,j}\}$ of rectangles will be a Markov partition
provided $(x,y)\in\interior{R_{i,j}}$ and $\phi(x,y)\in\interior{R_{i',j'}}$
imply that $\phi(W^u_{R_{i,j}}(x,y)) \supseteq W^u_{R_{i',j'}}(\phi(x,y))$
and $\phi(W^s_{R_{i,j}}(x,y)) \subseteq W^s_{R_{i',j'}}(\phi(x,y))$, where
$W^u_{R_{i,j}}(x,y)$ and $W^s_{R_{i,j}}(x,y)$ are the intersections with
$R_{i,j}$ of the unstable and stable manifolds, respectively, at $(x,y)$
\cite{Katok&Hasselblatt1995}.  For the baker map we have $W^u_{R_{i,j}}(x,y)
= [\frac{i-1}{M},\frac{i}{M}]\times\{y\}$ and $W^s_{R_{i,j}}(x,y) =
\{x\}\times[\frac{j-1}{N},\frac{j}{N}]$.  We consider the following three
cases separately:

\textbf{Case 1:\/} $\frac{i}{M}\le\half$

Since $x<\frac{i}{M}\le\half$, $(x',y')=\phi(x,y)=(2x,\frac{y}{2})$ and so
\begin{eqnarray*}
\phi\left(W^u_{R_{i,j}}(x,y)\right) &=& [\mathfrac{2i-2}{M},
\mathfrac{2i}{M}] \times \{\mathfrac{y}{2}\} \\
\phi\left(W^s_{R_{i,j}}(x,y)\right) &=& \{2x\} \times [\mathfrac{j-1}{2N},
\mathfrac{j}{2N}]
\end{eqnarray*}
and
\begin{eqnarray*}
W^u_{R_{i',j'}}(\phi(x,y)) &=& [\mathfrac{i'-1}{M}, \mathfrac{i'}{M}] \times
\{\mathfrac{y}{2}\} \\ W^s_{R_{i',j'}}(\phi(x,y)) &=& \{2x\} \times
[\mathfrac{j'-1}{N}, \mathfrac{j'}{N}]~.
\end{eqnarray*}
We now must determine $i',j'$ in terms of $i,j$.  Note that if $\frac{i-1}{M}<x<\frac{i}{M}-\frac{1}{2M}$ then $\frac{2i-2}{M}<x'<\frac{2i-1}{M}$, so $i'=2i-1$.  Similarly, if $\frac{i}{M}-\frac{1}{2M}\le x<\frac{i}{M}$ then $i'=2i$.  Note also that $\frac{j-1}{N}<y<\frac{j}{N}$ implies $\frac{j-1}{2N}<y'<\frac{j}{2n}$, so $j'=\frac{j}{2}$ if $j$ is even and $j'=\frac{j+1}{2}$ when $j$ is odd.  Thus,
\begin{eqnarray*}
x<\mathfrac{i}{M}-\mathfrac{1}{2M} &~:~& W^u_{R_{i',j'}}(\phi(x,y)) =
[\mathfrac{2i-2}{M}, \mathfrac{2i-1}{M}] \times \{\mathfrac{y}{2}\}~, \\
\mathfrac{i}{M}-\mathfrac{1}{2M}\le x &~:~& W^u_{R_{i',j'}}(\phi(x,y)) =
[\mathfrac{2i-1}{M}, \mathfrac{2i}{M}] \times \{\mathfrac{y}{2}\}~, \\
\mbox{$j$ even} &~:~& W^s_{R_{i',j'}}(\phi(x,y)) = \{2x\} \times
[\mathfrac{j-2}{2N}, \mathfrac{j}{2N}]~, \\
\mbox{$j$ odd} &~:~& W^s_{R_{i',j'}}(\phi(x,y)) = \{2x\} \times
[\mathfrac{j-1}{2N}, \mathfrac{j+1}{2N}]~.
\end{eqnarray*}

\textbf{Case 2:\/} $\frac{i-1}{M}\le\half<\frac{i}{M}$

For this case $M$ is necessarily odd and $i=\frac{M+1}{2}$.  We consider the
cases $x<\half$ and $x\ge\half$ separately.  For both cases we have \[
\phi\left(W^u_{R_{i,j}}(x,y)\right) = [1-\mathfrac{1}{M}, 1] \times
\{\mathfrac{y}{2}\} ~\cup~ [0,\mathfrac{1}{M}] \times
\{\mathfrac{y+1}{2}\}~. \]

\textbf{Case 2a:\/} $x<\half$

Since $x<\half$, $(x',y')=\phi(x,y)=(2x,\frac{y}{2})$.  From the fact that
$\frac{M-1}{2M}\le x<\half$ it follows that $\frac{M-1}{M}\le x'<1$, so
$i'=M$ and \[ W^u_{R_{i',j'}}(\phi(x,y)) = [\mathfrac{M-1}{M}, 1] \times
\{\mathfrac{y}{2}\}~. \]

For the stable manifold we have $\phi\left(W^s_{R_{i,j}}(x,y)\right) =
\{2x\} \times [\mathfrac{j-1}{2N}, \mathfrac{j}{2N}]$, so
\begin{eqnarray*}
\mbox{$j$ even} &~:~& W^s_{R_{i',j'}}(\phi(x,y)) = \{2x\} \times
[\mathfrac{j-2}{2N}, \mathfrac{j}{2N}]~, \\
\mbox{$j$ odd} &~:~& W^s_{R_{i',j'}}(\phi(x,y)) = \{2x\} \times
[\mathfrac{j-1}{2N}, \mathfrac{j}{2N}]~,
\end{eqnarray*}
since $j'=\frac{j}{2}$ when $j$ is even and $j'=\frac{j+1}{2}$ otherwise.

\textbf{Case 2b:\/} $\half\le x$

Since $x\ge\half$, $(x',y')=\phi(x,y)=(2x-1,\frac{y+1}{2})$.  From the fact
that $\half\le x<\frac{M+1}{2M}$ it follows that $0\le x'<\frac{1}{M}$, so
$i'=1$ and \[ W^u_{R_{i',j'}}(\phi(x,y)) = [0,\mathfrac{1}{M}] \times
\{\mathfrac{y+1}{2}\}~. \]

For the stable manifold we have $\phi\left(W^s_{R_{i,j}}(x,y)\right) =
\{2x-1\} \times [\mathfrac{j-1}{2N}+\half, \mathfrac{j}{2N}+\half]$.  From
the fact that $\frac{j-1}{N}<y<\frac{j}{N}$ it follows that
$\frac{j-1}{2N}+\half<y'<\frac{j}{2N}+\half$.  Thus, $j'=\frac{j+N}{2}$ if
$j+N$ is even, and $j'=\frac{j+N+1}{2}$ if $j+N$ is odd.  This gives us
\begin{eqnarray*}
\mbox{$j+N$ even} &~:~& W^s_{R_{i',j'}}(\phi(x,y)) = \{2x-1\} \times
[\mathfrac{j-2}{2N}+\half, \mathfrac{j}{2N}+\half]~, \\
\mbox{$j+N$ odd} &~:~& W^s_{R_{i',j'}}(\phi(x,y)) = \{2x-1\} \times
[\mathfrac{j-1}{2N}+\half, \mathfrac{j+1}{2N}+\half]~.
\end{eqnarray*}

\textbf{Case 3:\/} $\half<\frac{i-1}{M}$

Since $x>\frac{i}{M}>\half$, $(x',y')=\phi(x,y)=(2x-1,\frac{y+1}{2})$ and so
\begin{eqnarray*}
\phi\left(W^u_{R_{i,j}}(x,y)\right) &=& [\mathfrac{2i-2}{M}-1,
\mathfrac{2i}{M}-1] \times \{\mathfrac{y+1}{2}\} \\
\phi\left(W^s_{R_{i,j}}(x,y)\right) &=& \{2x-1\} \times
[\mathfrac{j-1}{2N}+\half, \mathfrac{j}{2N}+\half]
\end{eqnarray*}
and
\begin{eqnarray*}
W^u_{R_{i',j'}}(\phi(x,y)) &=& [\mathfrac{i'-1}{M}, \mathfrac{i'}{M}] \times
\{\mathfrac{y+1}{2}\} \\ W^s_{R_{i',j'}}(\phi(x,y)) &=& \{2x-1\} \times
[\mathfrac{j'-1}{N}, \mathfrac{j'}{N}]~.
\end{eqnarray*}
Note that if $\frac{i-1}{M}<x<\frac{i}{M}-\frac{1}{2M}$ then $\frac{2i-2}{M}-1<x'<\frac{2i-1}{M}-1$, so $i'=2i-1-M$.  Similarly, if $\frac{i}{M}-\frac{1}{2M}\le x<\frac{i}{M}$ then $i'=2i-M$.  Note also that $j'=\frac{j+N}{2}$ if $j+N$ is even and $j'=\frac{j+N+1}{2}$ when $j+N$ is odd.  Thus,
\begin{eqnarray*}
x<\mathfrac{i}{M}-\mathfrac{1}{2M} &~:~& W^u_{R_{i',j'}}(\phi(x,y)) =
[\mathfrac{2i-2}{M}-1, \mathfrac{2i-1}{M}-1] \times \{\mathfrac{y+1}{2}\}~,
\\
\mathfrac{i}{M}-\mathfrac{1}{2M}\le x &~:~& W^u_{R_{i',j'}}(\phi(x,y)) =
[\mathfrac{2i-1}{M}-1, \mathfrac{2i}{M}-1] \times \{\mathfrac{y+1}{2}\}~, \\
\mbox{$j+N$ even} &~:~& W^s_{R_{i',j'}}(\phi(x,y)) = \{2x-1\} \times
[\mathfrac{j-2}{2N}+\half, \mathfrac{j}{2N}+\half]~, \\
\mbox{$j+N$ odd} &~:~& W^s_{R_{i',j'}}(\phi(x,y)) = \{2x-1\} \times
[\mathfrac{j-1}{2N}+\half, \mathfrac{j+1}{2N}+\half]~.
\end{eqnarray*}

This completes the proof.
\qed
\end{pf}




\begin{thebibliography}{9}

\bibitem{Antoniou1997}
I. Antoniou and K. Gustafson,
From irreversible Markov semigroups to chaotic dynamics,
\emph{Physica A\/} \textbf{236} (1997) 296--308.

\bibitem{Arnold&Avez1968}
V. Arnold and A. Avez,
\emph{Ergodic Problems of Classical Mechanics\/}
(Benjamin, New York, 1968).

\bibitem{Cornfeld1982}
I. Cornfeld, S. Fomin, and Y. Sinai,
\emph{Ergodic Theory\/}
(Springer-Verlag, New York, 1982).

\bibitem{Cox&Miller1965}
D. Cox and H. Miller,
\emph{The Theory of Stochastic Processes\/}
(Metheun, London, 1965).

\bibitem{Davies1976}
E. Davies,
\emph{Quantum Theory of Open Systems\/}
(Academic Press, London, 1976).

\bibitem{Dudley1989}
R. Dudley,
\emph{Real Analysis and Probability\/}
(Chapman \& Hall, New York, 1989).

\bibitem{Ehrenfest1959}
P. Ehrenfest and T. Ehrenfest,
\emph{The Conceptual Foundations of the Statistical Approach in Mechanics\/}
(Cornell University Press, Ithaca, NY, 1959).

\bibitem{Hannay1994}
J. Hannay, J, Keating, A. Ozorio,
Optical realization of the baker's transformation,
\emph{Nonlinearity\/} \textbf{7} (1994) 1327--1342

\bibitem{Hasegawa1992}
H. Hasegawa and W. Saphir,
Non-equilibrium statistical mechanics of the baker map: Ruelle resonances and subdynamics,
\emph{Phys. Let. A\/} \textbf{161} (1992) 477--482.

\bibitem{Lasota1994}
A. Lasota and M. Mackey,
\emph{Chaos, Fractals, and Noise\/}
(Springer-Verlag, New York, 1994).

\bibitem{Nicolis1988}
G. Nicolis, C. Nicolis,
Master-equation approach to deterministic chaos,
\emph{Phys. Rev. A\/} \textbf{38} (1988) 427--433.

\bibitem{Kampen1962}
N. van Kampen,
Fundamental Problems in the Statistical Mechanics of Irreversible Processes,
in: E. Cohen ed.,
\emph{Fundamental Problems in Statistical Mechanics\/}
(North-Holland, Amsterdam, 1962).

\bibitem{Kampen1981}
N. van Kampen,
\emph{Stochastic Processes in Physics and Chemistry\/}
(North Holland, Amsterdam, 1981).

\bibitem{Katok&Hasselblatt1995}
A. Katok and B. Hasselblatt
\emph{Introduction to the Modern Theory of Dynamical Systems\/}
(Cambridge University Press, Cambridge, 1995).

\bibitem{Misra1980}
B. Misra and I. Prigogine,
On the foundations of kinetic theory,
\emph{Progr. Theor. Phys. Supp.\/} \textbf{69} (1980) 101--110.

\bibitem{Ross1983}
S. Ross,
\emph{Stochastic Processes\/}
(John Wiley \& Sons, New York, 1983).

\end{thebibliography}
\end{document}